\documentclass{sig-alternate}

\usepackage{multirow}

\usepackage{enumerate}
\usepackage{graphicx}
\usepackage{amsfonts}
\usepackage{amssymb}
\usepackage[caption=false]{subfig}
\usepackage{stmaryrd}
\usepackage{algorithm}
\usepackage{algorithmic}

\usepackage{caption}
\DeclareCaptionType{copyrightbox}
\usepackage{subfig}

\newtheorem{definition}{Definition}

\newcommand{\hide}[1]{}

\begin{document}

\title{Finding the Right Set of Users: Generalized Constraints for Group Recommendations}

\conferenceinfo{PersDB}{2012}

\numberofauthors{2}
\author{
\alignauthor
Kostas Stefanidis\\
       \affaddr{Dept. of Computer and Information Science}\\
       \affaddr{Norwegian University of Science and Technology, Norway}\\
       \email{kstef@idi.ntnu.no}
\alignauthor 
Evaggelia Pitoura\\
       \affaddr{Dept. of Computer Science}\\
       \affaddr{University of Ioannina, Greece}\\
       \email{pitoura@cs.uoi.gr}
}

\maketitle

\begin{abstract}
Recently, group recommendations have attracted considerable attention.
Rather than recommending items to individual users, group recommenders
recommend items to groups of users.
In this position paper, we introduce the problem of forming an appropriate group of users to recommend
an item when constraints  apply to the members of the group. 	
We present a formal model of the problem and an algorithm for its solution.
Finally, we identify several directions for future work.
\end{abstract}

\section{Introduction}
Recommendation systems aim at suggesting to users items of potential interest to them. 
In general, this is achieved by estimating a rating for each item and  user 
and then recommending to the user the item with the highest rating \cite{DBLP:journals/tkde/AdomaviciusT05}.
In the content-based approach (e.g.,\ \cite{DBLP:journals/ml/PazzaniB97,DBLP:conf/dl/MooneyR00}), 
the estimation of the rating of an item is based on the ratings that the user has assigned to
similar items, whereas   
in collaborative filtering systems (e.g.,\ \cite{DBLP:journals/cacm/KonstanMMHGR97,DBLP:conf/uai/BreeseHK98}),
this rating is predicted using previous ratings of the item by similar users.
Knowledge-based approaches enhance recommendations
by exploiting domain knowledge \cite{Burke2000}.
Most previous work focuses on recommending individual items to individual users.
However, recently, group recommendations have received considerable attention.
Instead of recommending items to individual users, group recommenders make 
recommendations to groups of 
users (e.g.,\ \cite{DBLP:journals/vldb/RoyACDY10,DBLP:conf/recsys/BaltrunasMR10,DBLP:conf/dasfaa/NtoutsiSNK12,DBLP:conf/group/GartrellXLBHMS10,springerlink:10.1007/978-3-642-16089-9_1}). 

An effective group recommendation system must take into account not only the 
preferences of individual users but also the group dynamics, 
that is, how groups of people make decisions. 
Since an item must be acceptable by all users of the group,
different consensus functions or strategies have been derived that 
characterize how much the item satisfies
the group as a whole.
For example,
group consensus may be estimated based on disagreement and relevance, where disagreement 
expresses the difference in the item ratings between 
the group members, while relevance   
corresponds, for example, to the average of the ratings of the item for the group members, 
or, to the highest  (in the optimistic strategy), 
or lowest (in the least misery strategy) rating among the ratings for the group members \cite{DBLP:journals/vldb/RoyACDY10}.

The ``reverse'' problem has also been studied, that is,  recommending groups or packages 
of items to individual users (e.g.,\ \cite{DBLP:conf/sigmod/RoyACDY10,DBLP:conf/recsys/XieLW10}). 
For example, in {\em top-k composite recommendations}, each recommendation consists of a  
set of items, where both an interest score or rating and a cost is associated with each item \cite{DBLP:conf/recsys/XieLW10}. 
The user specifies a maximum total cost for any recommended set.
Motivated by online shopping applications, recommending {\em composite items} is proposed in 
\cite{DBLP:conf/sigmod/RoyACDY10}, 
where a set of related satellite items are recommended along with each central item.
In addition, the problem of recommending sets of items to a user  
when the items to be recommended must satisfy several constraints
is considered in \cite{DBLP:journals/tois/ParameswaranVG11}.  
Particular focus is given on recommending 
sets of courses to students in the context of the CourseRank project.

In this paper, we consider a different aspect of group recommendations. 
Whereas in previous research, group recommendations focus on the relevance of the item to the group members, 
here, we study group recommendations when 
specific constraints apply to the members of the group based on individual preferences
that the members of the group express for the other group participants.
For example, a vacation package may seem more attractive to a user,
if the other members of the group are of a similar age. 
Furthermore, a course may be recommended to a group of students that have similar or diverse 
backgrounds depending on the scope of the course, whereas a user may prefer a recommendation for 
a specific restaurant, if the accompanying group members are non smokers.

In the rest of this paper, we present in some detail a specific motivating example.
Then, we introduce a formal model of the problem and outline a solution. We conclude the paper with
a list of issues for future work.

\section{Motivating Example}
Assume a travel agent web site promoting vacation packages. 
As an example, consider the package ``Gems of the Aegean'' 
referring to a cruise itinerary at Eastern Mediterranean with 
tagline: ``cities, sights and sensational scenery''. 
The cruise starts at July 16, 
its duration is 8 days, costs 900 euro and visits Athens, 
Kalamata, Aghios Nikolaos, Kusadasi and Marmaras.
The goal of such a travel agent company is to 
locate an appropriate group of people which will be highly interested in the cruise.

Each user in the system has expressed a set of constraints 
concerning his/her choices. 
These constraints refer to either the vacation package itself or the other members of 
the group that potentially he/she will be a member. 
For example, Alice, a 34 years old teacher planning her vacations, 
can spend up to 1000 euro for less than 10 days in July, 
is interested in seeing the cities and sights of 
Eastern Mediterranean, and prefers to be accompanied by people  
over thirty most of which are college graduates. 

Moreover, the company itself has defined a set of constraints 
expressing the preferences of the company for the group of customers 
that the company is targeting on. 
For instance, consider a constraint that directs the selection toward groups 
with middle-aged and senior people that is formulated as ``include at least 
30 users with average age greater than 40 years old''.

Given that for each user, there is an interest score 
available for each item, e.g., vacation package, that 
indicates how desirable the item is for the user, 
the task of the recommender is to recommend an item to a group of users, such that, 
(i) the (average) interest scores for all users in the group is maximized, and 
(ii) the constraints of all users in the group, as well as the constraints of the company, are satisfied.

\section{Model and Problem Statement}
In this section, we present a model for constraints that captures different kinds of features for user group construction. We distinguish between user-to-item, user-to-group and group-to-group constraints. Then, we introduce the problem of constructing groups of users satisfying a set of constraints.

\subsection{User-to-Item Constraints}
Let $t$ be an item, i.e., a vacation package in our example, described by a set of attributes 
$\{a_1, \ldots, a_p\}$, where each $a_i$, $1 \leq i \leq p$, is of the form ($a_i.attribute = a_i.value$).
The description of the cruise {\em Gems of the Aegean} is shown in Figure~\ref{fig:itemexample}.

\begin{figure}[h]
\scriptsize
\centering
	\begin{tabular}{|p{0.5in}p{0.1in}p{1.5in}|}
	\hline
	title & = & Gems of the Aegean \\
	type & = & cruise \\
	place & = & Eastern Mediterranean \\
	locations & = & Athens, Kalamata, Aghios Nikolaos, Kusadasi, Marmaras \\ 
	start\_date & = & July 16, 2012 \\
	end\_date & = & July 24, 2012 \\
	duration & = & 8 \\
	cost & = & 900 \\
	\hline
	\end{tabular}
\caption{Item description example.}
\label{fig:itemexample}
\end{figure}

Let $U$ be a set of users.
We use $score(u,t)$, $u \in U$, to denote the relevance, value or importance of item $t$ for user $u$.
The value of $t$ for $u$ can, for example, be directly induced by preferences that $u$ has expressed in her profile. 
For example, given that Alice enjoys sightseeing in 
warm climates and likes to travel by sea, the {\em Gems of the Aegean} cruise
will receive a high preference score.
When there are no explicit preferences, $score(u,t)$ may be the predicted rating of a recommender 
based on the past behavior of $u$ or of other similar users.

In addition, each user $u$ may specify a set of basic constraints $\{b_1, \ldots, b_q\}$ on the values of specific attributes of items that are candidates to be recommended. 
Each attribute constraint has an attribute name, a binary operator and a value, i.e., has the form ($b_i.attribute$ $\theta_{b_i}$ $b_i.value$). Attributes and values have the same form as in items. Binary operators include common operators, such as $=$, $\neq$, $<$, $\leq$, $>$, $\geq$ and {\em substring}.
We refer to such constraints as {\em user-to-item constraints}.
Figure~\ref{fig:u-to-iexample} depicts the user-to-item constraints of Alice.

\begin{figure}[h]
\scriptsize
\centering
	\begin{tabular}{|p{0.6in}p{0.1in}p{1.5in}|}
	\hline
	place & = & Eastern Mediterranean \\
	start\_ date & $\geq$ & July 1, 2012 \\
	end\_ date & $\leq$ & July 31, 2012 \\
	duration & $<$ & 10 \\
	cost & $\leq$ & 1000 \\
	\hline
	\end{tabular}
\caption{User-to-item constraints example.}
\label{fig:u-to-iexample}
\end{figure}

Such user constraints are fulfilled with respect to an item $t$, if and only if, every attribute constraint of the user is satisfied by some attribute of $t$. Formally:

\begin{definition}[User-to-Item Constraints]
An item $t$ is a set \{$a_1, \ldots, a_p$\}, where 
each $a_i$, $1 \leq i \leq p$, is of the form ($a_i.attribute = a_i.value$). 
The user-to-item constraints of user $u$ is a set  \{$b_1,$ $\ldots,$ $b_q$\}, 
where each $b_j$, $1 \leq j \leq q$, is of the form ($b_j.attribute$ $\theta_{b_j}$ $b_j.value$) and $\theta_{b_j} \in \{=, \neq, <, \leq, >, \geq,~substring\}$.

$t$ satisfies the user-to-item constraints of $u$, if and only if, $\forall$ $b_j$, $\exists$ $a_i$, such that, $a_i.attribute$ $=$ $b_j.attribute$ and $((a_i.value)$ $\theta_{b_j}$ $(b_j.value))$, $1 \leq i \leq p$, $1 \leq j \leq q$.
\end{definition}

For example, the package of Figure~\ref{fig:itemexample} satisfies the user-to-item constraints of Figure~\ref{fig:u-to-iexample}, which means that the personal constraints of Alice for the cruise are fulfilled.

\subsection{User-to-Group Constraints}
Our overall goal is to form a group of users for which the recommended item is highly valuable for the group as a whole. 
Apart from the user-to-item constraints, we are also interested in satisfying user requirements that refer to user choices concerning the other members of the group. The fulfilling of these requirements works towards ensuring low disagreement 
between the group members. 
We refer to such constraints as {\em user-to-group constraints}.

In particular, similar to items, we assume that each user $u \in U$ is described by a set of attributes $\{x_1, \ldots, x_s\}$, where each $x_i$, $1 \leq i \leq s$, is of the form ($x_i.attribute = x_i.value$). For example, an attribute can be {\em name}, {\em education}, {\em occupation}, {\em gender} or {\em age}. See, for instance, Figure~\ref{fig:userexample} for a user description.

\begin{figure}[h]
\scriptsize
\centering
	\begin{tabular}{|p{0.7in}p{0.1in}p{1.0in}|}
	\hline
	name & = & Alice \\
	education & = & college graduate \\
	occupation & = & educator \\
	gender & = & female \\
	age & = & 34 \\
	\hline
	\end{tabular}
\caption{User description example.}
\label{fig:userexample}
\end{figure}

Each user defines a set of basic constraints $\{y_1, \ldots, y_r\}$ on the values of specific attributes of the other users 
belonging to his/her group.
For example, consider the preference of Alice in groups with members over thirty, i.e., $age > 30$.
This kind of constraints, called {\em value constraints}, are similar to the user-to-item constraints.

More complex constraints that express aggregation requirements
on the values of the set, termed {\em aggregation constraints}, are also permitted.
In particular, the constraint $z_i$ defined as:\\ 
($aggr_{\mathcal{G}}(z_i.attribute)$ $\theta_{z_i}$ $z_i.value$), 
where $\theta_{z_i} \in$ $\{=$, $<$, $\leq$, $>$, $\geq\}$ and $aggr_{\mathcal{G}}$ is either $avg$, $sum$, $max$, $min$
or $count$
is a constraint respectively on the average, sum, maximum, minimum or 
number of the values of $z_i.attribute$ of group $\mathcal{G}$.
For example, take Scott, a 20 years college student. The constraint $avg_{\mathcal{G}}(age) < 25$ of 
Scott means that he is interested in groups with users with average age smaller than 25 years old.

Value and aggregation constraints apply to all members of the
group. Often, we want to specify constraints on subsets of the group.
To this end, we define {\em composite constraint} $w_i$ of the form:\\
$include$ $at$ $least$ $l$ $users$ $with$ ($w_i.attribute$ $\theta_{w_i}$ $w_i.value$), or\\
$include$ $at$ $least$ $l$ $users$ $with$ ($aggr_{\mathcal{G}}(w_i.attribute)$ $\theta_{w_i}$\\ $w_i.value$).\\
For example, $include$ $at$ $least$ $10$ $users$ $with$ ($age$ $>$ $35$), expresses the preference 
of a user in groups with at least 10 users with age greater than 35 years old.
Clearly, we could also consider conjunction of constraints. An example such constraint
could be for groups that include at least 10 women of average age below twenty.

Given a group $\mathcal{G}$ of users and a user $u$, 
we say that the user-to-group constraints of $u$ are fulfilled with respect to $\mathcal{G}$, if and only if, every value constraint of $u$ is satisfied by every user in $\mathcal{G}$ and every aggregation and composite constraint of $u$ is satisfied by the members of $\mathcal{G}$ as a whole.

The following definition formalizes our model of user-to-group constraints.

\begin{definition}[User-to-Group Constraints]
Let  $u$ be a user described by a set of attributes $\{x_1, \ldots, x_s\}$, where each $x_i$, $1 \leq i \leq s$, 
is of the form ($x_i.attribute = x_i.value$), and $\mathcal{G}$ be a set of users. 
$\mathcal{G}$ satisfies the user-to-group constraints of $u$, if and only if:
\begin{enumerate}
	\item[(i)] [$\mathcal{G}$ satisfies the value constraints of $u$] For each value constraint $y$ of the form ($y.attribute$ $\theta_y$ $y.value$) of $u$, $\theta_y \in \{=, \neq, <, \leq, >, \geq,~substring\}$, there exists, for each user in $\mathcal{G}$, a $x$ attribute, such that, $x.attribute = y.attribute$ and $((x.value)$ $\theta_{y}$ $(y.value))$.
	\item[(ii)] [$\mathcal{G}$ satisfies the aggregation constraints of $u$] For each aggregation constraint $z$ of $u$,
it holds $aggr_{\mathcal{G}}(z.attribute)$ $\theta_{z}$ $z.value$, where $\theta_z \in$ $\{=$, $<$, $\leq$, $>$, $\geq\}$ and $aggr_{\mathcal{G}}$ is $avg$, $sum$,  $count$, $min$ or $max$.
	\item[(iii)] [$\mathcal{G}$ satisfies the composite constraints of $u$] For each composite constraint ``$include$ $at$ $least$ $l$ $users$ $with$ $w$'' of $u$, there exist at least $l$ users in $\mathcal{G}$ forming a group $\mathcal{G}'$, 
$\mathcal{G}'$ $\subseteq$ $\mathcal{G}$ and $|\mathcal{G}'|$= $l$,
such that, $\mathcal{G}'$ satisfies the value or aggregation constraint $w$ of $u$.
\end{enumerate}
\end{definition}

\subsection{Group-to-Group Constraints}
Both user-to-item and user-to-group constraints describe limitations from the user, or customer, perspective. From the perspective of the company, {\em group-to-group constraints} refer to a set of properties that the group under construction  
must satisfy. 
These properties express the requirements of the company 
concerning the group that a product, or item, is targeting on.

As with the user-to-group constraints, we distinguish group-to-group constraints into value, aggregation and composite ones.
For example, assume that the travel agent company managing the package {\em Gems of the Aegean} targets mainly on 
college graduates or say, middle-aged and senior customers. 
In this case, the company aims at a group $\mathcal{G}$ that should qualify, for instance, the composite constraint {\em include at least 20 users with $education$ = `college graduate'} or 
the composite constraint {\em include at least 20 users with $avg_{\mathcal{G}}(age)>40$}.
When such constraints are fulfilled for a group $\mathcal{G}$, we say that $\mathcal{G}$ {\em satisfies the group-to-group constraints}.
(The formal definition of group-to-group constraints is skipped, since it is similar to the definition of user-to-group constraints.)

\subsection{Problem Statement}
A group $\mathcal{G}$ is called {\em satisfiable} if all the {\em user-to-item}, {\em user-to-group} and {\em group-to-group} constraints for $\mathcal{G}$ are fulfilled with respect to an item $t$.

\begin{definition}[Satisfiable Group]
Given an item $t$ and a group $\mathcal{G}$, $\mathcal{G}$ is satisfiable for $t$, if and only if:
\begin{enumerate}
\item[(i)] $t$ satisfies the user-to-item constraints of $u$, $\forall u \in \mathcal{G}$,
\item[(ii)] $\mathcal{G}$ satisfies the user-to-group constraints of $u$, $\forall u \in \mathcal{G}$, and
\item[(iii)] $\mathcal{G}$ satisfies the group-to-group constraints.
\end{enumerate}
\end{definition}

Next, we formally define the problem of finding an appropriate 
group of users with the maximum relevance score for a specific item.

\begin{definition}[Problem Definition]
Given an item $t$, a set of users $U$ and an integer $k$, identify the group of users $\mathcal{G}$, $\mathcal{G} \subseteq U$, with cardinality $k$, such that:
\begin{enumerate}
	\item[(i)] $\mathcal{G}$ is satisfiable for $t$ and
	\item[(ii)] $\mathcal{G}$ has the maximum value $score({\mathcal{G}, t})$, where 
	\[score({\mathcal{G}, t}) = \sum_{u \in \mathcal{G}}(score(u, t)),\]
	among all satisfiable groups with cardinality $k$.
\end{enumerate}
\end{definition}

In this paper, we assume that constraints are strong, in the sense, that 
for recommending an item to a group all related constraints must be satisfied. 
One could also envision models where users prioritized their constraints, or where
one type of constraint is more important than another, for instance,
user-to-group constraints may be more relevant than group-to-group constraints.
Further, we assume that it is possible to form satisfiable groups.
Clearly, in the general case, it may not be possible to construct  
such  groups.  In this case, we should look for approximate solutions,  
for example, for a group that satisfies the largest number of constraints or
for a group that satisfies the constraints of the majority of its members.

\section{A Greedy Algorithm}
Given an item $t$, a satisfiable group $\mathcal{G}$ for $t$ is a set of users that 
comply with a set of user-to-item, user-to-group and group-to-group constraints. 
For this specific item $t$, a technical challenge is to efficiently construct 
the satisfiable group $\mathcal{G}$ with the highest value $score(\mathcal{G}, t)$, given a budget $k$ on the size of $\mathcal{G}$.

A brute-force method to identify such a 
group is to first construct all combinations of $k$ users forming a satisfiable group $\mathcal{G}$ and then pick the one with the maximum $score(\mathcal{G}, t)$. 
Constructing the group $\mathcal{G}$ using such a straight-forward algorithm is computational costly, since the number of groups to examine for satisfyingness can be overwhelming even for a small number of users. 
As a result, we propose an alternative algorithm that computes an approximate group $\mathcal{G}$.

In particular, we use the following intuitive heuristic. 
We incrementally construct a group of users by selecting at each step a user that: (i) adds the most to the score value of the group and 
(ii) after joining the group of users already selected, the group satisfies more constraints than after adding any other user.

More specifically, let $t$ be an item, $U$ be a set of users and $\mathcal{G}$ be the set we want to construct.
Let also $\mathcal{A}$ be the users in $U$, such that, $\forall u \in \mathcal{A}$, $t$ satisfies the user-to-item constraints of $u$.
Initially, $\mathcal{G}$ is empty. 
We first construct a single-user group by adding  to $\mathcal{G}$ a user $u$, $u \in \mathcal{A}$, 
randomly selected among the ones with the maximum value $score(u, t)$.
Then, at each step, we select the users in $\mathcal{A} \backslash \mathcal{G}$ with the maximum value score for $t$, and add to $\mathcal{G}$ the one that satisfies along with the other 
members of $\mathcal{G}$ the most user-to-group and group-to-group constraints.
This procedure stops after generating a group $\mathcal{G}$ with $k$ users.
Note that we assume that there is at least one group, such that, 
item $t$ satisfies the {\em user-to-item} constraints of at least $k$ users.   
Algorithm~\ref{alg:group} illustrates our {\em Group Construction Algorithm}.

\begin{algorithm}[h]
\small
\caption{Group Construction Algorithm}
\label{alg:group}
\begin{algorithmic}[1]
  \REQUIRE{An item $t$, a set of users $U$ and an integer $k$.}
  \ENSURE{A group $\mathcal{G}$, $\mathcal{G} \subseteq U$, of $k$ users.}
  \vspace{0.1cm}
  \hrule
  \vspace{0.1cm}
  \STATE \textbf{begin}
  \STATE $\mathcal{G} \leftarrow \emptyset$;
  \STATE find the users $\mathcal{A}$, $\mathcal{A} \subseteq U$, such that, $\forall u \in \mathcal{A}$, $t$ satisfies the user-to-item constraints of $u$;
  \STATE find the users $\mathcal{B}$, $\mathcal{B} \subseteq \mathcal{A}$, with the maximum value score for $t$;
  \STATE randomly select a user $u \in \mathcal{B}$;
  \STATE $\mathcal{G} \leftarrow \mathcal{G} \cup \{u\}$;
 
  \WHILE{$|\mathcal{G}| < k$} 
    \STATE find the users $\mathcal{C}$, $\mathcal{C} \subseteq \mathcal{A} \backslash \mathcal{G}$, with the maximum value score for $t$;
    \STATE select the user $u_{add} \in \mathcal{C}$, such that, $\mathcal{G} \cup \{u_{add}\}$ satisfies the most user-to-group and group-to-group constraints compared to the groups $\mathcal{G} \cup \{u\}$, $\forall u \in \mathcal{C}$;
    \STATE $\mathcal{G} \leftarrow \mathcal{G} \cup \{u_{add}\}$;
  \ENDWHILE

  \RETURN $\mathcal{G}$;               
  \STATE \textbf{end}       
\end{algorithmic}
\end{algorithm}

We illustrate the final step of this algorithm with the following example. Consider the package {\em Gems of the Aegean} and a travel agent company that wants to construct a group of 80 users. At the last step, the 79 users have already been selected. Then, the users, say Alice and Scott, with the maximum interest score for the cruise, say 0.9, are identified. 
The cruise satisfies the user-to-item constraints of both users.
Based on the assumption that when adding Alice to the group more user-to-group and group-to-group constraints are fulfilled than when adding Scott, Alice is the last user that will join the group.

Note that in line 9 of the algorithm, we are interested in locating the user that when joining a group, the group fulfills more constraints or, in other words, violates less constraints than when a different user joins the group. Several meanings can be assigned to the expression ``{\em fulfills more}'' or, respectively, ``{\em violates less}'' constraints.
For instance, assume a value user-to-group constraint expressing a preference of a user $u$. Assume also the following two scenarios: (i) all users in a group violate the value constraint of $u$ or (ii) all users, except one, fulfill this constraint. It is clear that in both cases the group does not satisfy the constraint of $u$. 
However, different policies for counting the number of the violated constraints can be applied. For this example, for instance, we can consider a scheme that assigns a weight to each constraint depending on the number of the non-satisfied users, instead of considering each constraint as being either satisfiable or not.

\section{Discussion and Future Work}
In this short position paper, we introduced the generalized group formation 
problem: how to form a group of users 
for recommending an item, such that, all members of the group are satisfied both from 
the selection of the recommended item and from the selection of the other group members. 
Whereas the first component of the model, that is, group consensus on an item, 
has received considerable attention, the other dimension of the problem, 
that is, group consensus on the other group members is, to our knowledge, novel.

In this paper, we outlined a model for the problem and presented a first greedy algorithm for
its solution. Clearly, there are many directions for future work including modeling constraints, 
combining item and group preferences, efficient algorithms and finally implementations in specific contexts. 
Next, we elaborate on these issues a bit further.

We modeled user preferences on items and other group members as constraints on the values of the attributes 
of the item and the user respectively. Other models are feasible as well. In particular, the user-to-item 
part can be expressed using traditional content or collaborative filtering approaches and be replaced 
by a single relevance value. 
One can envision a similar approach to expressing the user-to-group part. Recommendation 
techniques could also be used to recommend group members that would be compatible with a specific user,
for example, similar users or users that liked similar items in the past.

In many cases, a group that satisfies all user-to-item and user-to-group constraints may not exist.
In such cases, an appropriate consensus function should be sought for. For instance, one
may ask for a group that satisfies the largest number of its members. 
Other approaches, such as relaxing constraints or approximating them, are also feasible.
Further, an item may be highly relevant to many users but these users may be incompatible with each other.
In this case, it may be wise to recommend this item to a less relevant but more compatible group. 

Moreover, depending on the type and complexity of 
the constraints, more efficient algorithms than the greedy algorithm can be designed.
Finally, in this paper, we used travelling as an example. Group constraints apply in many areas, where people
perform activities in groups. For example, one could explore them  in conjunction with social networks.

\section{Acknowledgments}
The work of the first author was carried out during the tenure of an ERCIM ``Alain Bensoussan'' Fellowship Programme. This Programme is supported by the Marie Curie Co-funding of Regional, National and International Programmes (COFUND) of the European Commission.
The work of the second author is partially supported  
by project ``Intersocial'', financed by the European Territorial Cooperation Operational Programme
"Greece-Italy" 2007-2013, co-funded by the European Union (European Regional
Development Fund) and by National Funds of Greece and Italy.

\bibliographystyle{abbrv}
\bibliography{persdb12_ref}

\end{document}